\documentclass[aps,prd,reprint,nofootinbib,superscriptaddress,longbibliography,floatfix,preprintnumbers]{revtex4-2}
\usepackage[T1]{fontenc}
\usepackage{amsmath,amssymb,bm,graphicx,booktabs,xcolor}
\usepackage[colorlinks=true,citecolor=blue,linkcolor=blue,urlcolor=blue]{hyperref}
\newcommand{\lcdm}{\ensuremath{\Lambda\mathrm{CDM}}}

\newcommand{\Dzero}{\ensuremath{\mathcal D_0}}
\newcommand{\Done}{\ensuremath{\mathcal D_1}}
\newcommand{\Mpl}{M_{\rm Pl}}
\newcommand{\dd}{\mathrm d}

\newcommand{\BFFZero}{0.63\pm0.61}

\newcommand{\BFAgZero}{1.87\pm0.61}

\newcommand{\BFQuiZero}{0.97\pm0.61}

\newcommand{\BFFOne}{2.19\pm0.61}
\newcommand{\RAWFOne}{2.21\pm0.61}

\newcommand{\BFAgOne}{1.17\pm0.61}
\newcommand{\RAWAgOne}{1.54\pm0.61}

\newcommand{\BFQuiOne}{1.79\pm0.61}
\newcommand{\RAWQuiOne}{2.28\pm0.60}

\hypersetup{pdftitle={Minimal vacuum histories beyond Lambda-CDM: model comparison within VCDM},pdfauthor={Antonio De Felice; Eleonora Di Valentino; Luis Escamilla Torres; Jack Zidich},pdfsubject={Coauthor draft: operationally converged MH annealing minima, 9 November 2026}}
\begin{document}
\preprint{YITP-26-160}
\title{Minimal vacuum histories beyond \texorpdfstring{$\Lambda$}{Lambda}CDM: model comparison within VCDM}
\author{Antonio De Felice}
\affiliation{Center for Gravitational Physics and Quantum Information,
Yukawa Institute for Theoretical Physics, Kyoto University, 606-8502, Kyoto, Japan}

\author{Eleonora Di Valentino}
\affiliation{School of Mathematical and Physical Sciences, University of Sheffield, Sheffield S3 7RH, United Kingdom}

\author{Luis A. Escamilla}
\affiliation{Department of Physics, Istanbul Technical University, Maslak 34469 Istanbul, Turkey}

\author{Jack Zidich}
\affiliation{Center for Gravitational Physics and Quantum Information,
Yukawa Institute for Theoretical Physics, Kyoto University, 606-8502, Kyoto, Japan}

\date{\today}
\begin{abstract}
Recent cosmological observations motivate alternatives to
$\Lambda$CDM with evolving dark energy. We compare three
such models, each introducing one additional cosmological
parameter: F42, with equation of state $w(a)=w_0/\sqrt{a}$,
and two smooth sign-changing effective dark-energy
profiles, Ag and Qui. All models are implemented within
VCDM, which determines the background and linear
perturbations without introducing an additional
propagating gravitational degree of freedom.
We use CMB observations from SPT, Planck, and ACT,
combined with DESI DR2 distance measurements, with
and without the DESY5 Dovekie supernova sample.
All three alternatives improve the best-fit likelihood
relative to $\Lambda$CDM. When supernovae are included,
the Akaike information criterion favors F42, while
Ag and Qui remain nearly indistinguishable from
$\Lambda$CDM on this criterion. The Bayesian evidence,
however, does not establish a robust preference among
the three alternatives. Our results
show that improved fits to current observations can
arise from physically different dark-energy histories,
motivating complementary measurements of expansion
and structure growth to distinguish them.
\end{abstract}

\maketitle

\section{Introduction}

The $\Lambda$ cold dark matter ($\Lambda$CDM) model provides a
simple description of the evolution of the Universe, in which
cold dark matter drives structure formation and a cosmological
constant accounts for the late-time accelerated expansion.
Within general relativity, the cosmological constant has a
constant energy density and equation of state $w=-1$.
Despite its simplicity, this framework has successfully described
a wide range of cosmological observations, including cosmic
microwave background (CMB) anisotropies~\cite{Planck:2018nkj,Planck:2018vyg,AtacamaCosmologyTelescope:2025blo,SPT-3G:2025bzu},
baryon acoustic oscillations (BAO)~\cite{DESI:2024mwx,DESI:2025zgx},
and Type Ia supernova (SN Ia)
distances~\cite{Scolnic:2021amr,DES:2024jxu}.
However, the physical nature of its dark components remains
unknown, and increasingly precise observations provide new
opportunities to test its assumptions.

Recent measurements from the Dark Energy Spectroscopic
Instrument (DESI), combined with CMB and SN Ia observations,
provide evidence that dark energy may evolve with
time~\cite{DESI:2025zgx,DESI:2025fii,DES:2025sig,Hoyt:2026fve}.
This preference persists after the Dovekie recalibration of
the DES five-year supernova sample and the updated treatment
of host-galaxy properties in Union3.1 and PantheonPlus.
With these updated supernova analyses, the combined constraints
give a $3.2$--$3.4\sigma$ preference over a cosmological constant
within flat $w_0w_a$CDM~\cite{DES:2025sig,Hoyt:2026fve}.
In this framework, the dark-energy equation of state follows
the Chevallier--Polarski--Linder (CPL) parameterization,
$w(a)=w_0+w_a(1-a)$~\cite{Chevallier:2000qy,Linder:2002et}.
The preferred evolution crosses the phantom divide, from
$w<-1$ at higher redshifts to $w>-1$ at the present epoch.
Evidence for this behavior has also been investigated beyond
the CPL parameterization, using alternative parameterizations
and nonparametric reconstructions~\cite{DESI:2025fii,Ozulker:2025ehg}.

These findings have motivated extensive investigations of
dynamical dark energy, exploring different cosmological datasets,
equation-of-state parameterizations, scalar-field models,
interacting dark sectors, and modifications of
gravity~\cite{Giare:2024ytc,DESI:2024mwx,Tada:2024znt,Luongo:2024fww,Cortes:2024lgw,Carloni:2024zpl,Wang:2024dka,Chan-GyungPark:2024mlx,Shlivko:2024llw,DESI:2024kob,Bhattacharya:2024hep,Gialamas:2024lyw,Najafi:2024qzm,Ye:2024ywg,Giare:2024gpk,Jiang:2024xnu,Reboucas:2024smm,RoyChoudhury:2024wri,Giare:2024oil,Wolf:2025jlc,Giare:2025pzu,DESI:2025zgx,DESI:2025fii,Kessler:2025kju,Santos:2025wiv,RoyChoudhury:2025dhe,Specogna:2025guo,Scherer:2025esj,Cheng:2025lod,Cheng:2025hug,Herold:2025hkb,Ozulker:2025ehg,Lee:2025pzo,Ishak:2025cay,Fazzari:2025lzd,Smith:2025icl,Li:2025vuh,Cheng:2025yue,Gokcen:2026pkq,Najafi:2026kxs,Yang:2026yaq,Kessler:2026dbi,Lee:2026yzs,Li:2026asg,Giare:2026oti,GuptaChoudhury:2026gsl,Giare:2026opg}.
Physically different mechanisms can reproduce similar expansion
histories, while predicting different signatures in cosmological
perturbations. The observations therefore invite a more targeted
question than whether \emph{some} parameterization of dark energy
can fit better: which physically distinct, yet comparably simple,
histories are supported by the data, and can the observations
distinguish them?

A useful first step is to restrict the additional cosmological
freedom. Here we compare $\Lambda$CDM with three histories requiring
only one additional sampled parameter relative to $\Lambda$CDM.
The first is F42, a smooth evolving equation of state
$w(a)=w_0/\sqrt{a}$ capable of phantom
crossing~\cite{Borghetto:2026ytv}, where $a$ is the scale
factor normalized to unity today.
The other two are smooth sign-changing effective dark-energy
histories, denoted Ag (agitated) and Qui (quiescent), with a
characteristic transition scale $a_\dagger$~\cite{Akarsu:2024qsi,Akarsu:2024eoo}.
The Ag profile includes a localized density enhancement around
the transition, while the Qui profile adopted here is monotonic.
These choices are not intended to span every dynamical-dark-energy
model. Rather, they illustrate two qualitatively different
histories---continuous evolution of a positive effective
dark-energy density and a transition between negative and
positive effective dark-energy density---at matched parameter count.
Their precise forms are given in Sec.~\ref{sec:models}.

Phenomenological freedom alone does not specify a theory of
perturbations. We consequently implement all four backgrounds within
VCDM, a type-II minimally modified gravity theory with a Lagrangian
formulation and the same two propagating gravitational degrees of
freedom as general relativity~\cite{DeFelice:2020eju,DeFelice:2022uxv}.
Its auxiliary potential can be reconstructed for the chosen
expansion history, while its field equations determine cosmological
perturbations. The single \emph{extra parameter} in each background
should not be confused with an extra propagating field.
This common framework permits the background and perturbative
predictions to be confronted with data without introducing an
arbitrary dark-energy sound speed.
It is a consistent realization of the chosen histories, not a claim
that VCDM is their only possible theoretical completion.

In this work, we compare the observational support for these
minimal histories and investigate whether current cosmological
measurements can distinguish them. Our results show that
different histories can improve the fit relative to $\Lambda$CDM,
without establishing a unique explanation of cosmic acceleration.
This motivates complementary measurements of expansion and
structure growth, notably those targeted by
Euclid~\cite{Euclid:2024yrr}.

The paper is structured as follows.
Section~\ref{sec:models} introduces the theoretical models,
while Secs.~\ref{sec:data} and~\ref{sec:evidence} describe
the datasets and statistical framework.
Section~\ref{sec:results} presents the results, followed by
their physical interpretation and prospects for new data
in Sec.~\ref{sec:limits}.
Further numerical details and the complementary Cobaya
MCMC analysis are provided in the appendices.

\section{Models and common assumptions}\label{sec:models}

We assume a spatially flat Universe with adiabatic scalar initial
conditions and the same matter and radiation content for all models.
The neutrino sector consists of three degenerate massive species
with a fixed total mass $\sum m_\nu=0.06\,\mathrm{eV}$ and
temperatures corresponding to $N_{\rm eff}=3.044$.
The primordial helium abundance is determined from BBN
consistency using a PArthENoPE-based interpolation table.
We adopt a pivot scale $k_*=0.05\,\mathrm{Mpc}^{-1}$ for the
primordial scalar power spectrum.

\subsection{The role of VCDM}

VCDM is a type-II minimally modified gravity theory that
propagates two gravitational degrees of freedom, as in general
relativity, while an auxiliary potential can realize a chosen
homogeneous cosmological history~\cite{DeFelice:2020eju,DeFelice:2022uxv}.
The effective density and pressure introduced below describe
the background implied by the gravitational theory, rather
than a separately added dark-energy fluid.

The homogeneous equations may be written in terms of the
auxiliary field $\phi$ and potential $V(\phi)$
as~\cite{DeFelice:2020eju}
\begin{align}
 H&=\frac12 V_{,\phi}-\frac{\phi}{3},\label{eq:vcdm}\\
 \dot\phi&=\frac{3(\rho+p)}{2\Mpl^2},\\
 3\Mpl^2H^2&=\rho+\rho_{\rm DE},\\
 \rho_{\rm DE}&=\Mpl^2\left(V-\phi V_{,\phi}
                 +\frac34 V_{,\phi}^2\right).
\end{align}
Here $\rho$ and $p$ denote the total energy density and pressure
of matter and radiation, $\Mpl$ is the reduced Planck mass,
$V_{,\phi}\equiv\dd V/\dd\phi$, and a dot denotes a derivative
with respect to cosmic time.
Once the background is prescribed, these relations allow the
potential to be reconstructed, subject to regularity and
solution-branch requirements.
The VCDM field equations also determine the linear perturbations
without introducing a phenomenological dark-energy sound speed.
Agreement with a background expansion history in general
relativity therefore need not imply identical perturbations.

We calculate the linear CMB and matter-perturbation observables
using a modified version of CLASS~\cite{Lesgourgues:2011re,Blas:2011rf},
which implements the VCDM background and linear perturbation
equations. We use HMcode-2020~\cite{Mead:2020vgs} for nonlinear
corrections and adopt the same numerical accuracy settings
for all four models. Using the same prescription does not,
however, establish its validity for nonlinear VCDM structure
formation.

\subsection{F42}

The one-parameter F42 background is~\cite{Borghetto:2026ytv}
\begin{align}
 w(a)&=\frac{w_0}{\sqrt a},\label{eq:f42w}\\
 R_{\rm F42}(a)&\equiv\frac{\rho_{\rm DE}(a)}{\rho_{{\rm DE},0}}
 =a^{-3}\exp\left[6w_0\left(a^{-1/2}-1\right)\right].\label{eq:f42rho}
\end{align}
The density evolution in Eq.~\eqref{eq:f42rho} follows from
the continuity equation and is used in our implementation.
For $-1<w_0<0$, the phantom crossing occurs in the past, at
\begin{equation}
 a_\times=w_0^2,\qquad z_\times=w_0^{-2}-1.\label{eq:crossing}
\end{equation}
We use a uniform prior $w_0\in[-2,0]$.
No single value of $w_0$ makes Eq.~\eqref{eq:f42w} equal to
$-1$ at every epoch, so \lcdm\ is not a nested point of this
one-parameter family. Consequently, Wilks' theorem cannot
be used to interpret the likelihood improvement relative to
\lcdm\ as a significance with one degree of freedom.
The prior and data combination also differ from those of
Ref.~\cite{Borghetto:2026ytv}, so the corresponding Bayes
factors are not directly comparable.

\subsection{Ag and Qui sign-changing profiles}

We describe the Ag and Qui profiles through the normalized
effective dark-energy density,
\begin{align}
 R_B(a)&\equiv\frac{\rho_{\rm DE}(a)}{\rho_{{\rm DE},0}}
 =\frac{G_B(a)}{G_B(1)}, &
 G_B(a)&=\tanh x+B e^{-x^2},\label{eq:lambda-profile}\\
 x&=b\frac{a-a_\dagger}{a_\dagger}, & b&=10^{1.5}.
\end{align}
Ag has $B=1$, giving a localized positive bump around the
transition, while Qui has $B=0$, giving a monotonic hyperbolic
tangent. Both histories are smooth, with a characteristic
fractional transition width of order $b^{-1}$.
The transition scale $a_\dagger$ is the sole additional sampled
parameter, with a uniform prior $a_\dagger\in[0.25,0.99]$.
Allowing $b$ to vary would provide a more flexible shape,
but would also define a different model and prior volume
for evidence comparisons.

We define $z_\dagger=a_\dagger^{-1}-1$ as the redshift
corresponding to the center of the profile.
For Qui, it is also the redshift at which the density vanishes.
For Ag, the Gaussian contribution shifts the zero, so
$z_\dagger$ does not represent the sign-crossing redshift
for both models. Our comparison is restricted to the
profiles in Eq.~\eqref{eq:lambda-profile}.

The associated pressure follows from the continuity equation,
\begin{equation}
 p_{\rm DE}=-\rho_{\rm DE}-\frac13\frac{\dd\rho_{\rm DE}}{\dd\ln a}.
\end{equation}
We use the same density--pressure description in the
nonlinear-background interface. This avoids evaluating
$w=p_{\rm DE}/\rho_{\rm DE}$ at the density crossing,
where this ratio diverges even though the pressure remains finite.

\subsection{Earlier transitions and the \texorpdfstring{\lcdm}{Lambda-CDM} limit}\label{sec:adaggerlimit}

The early-transition limit is relevant when interpreting posterior
probability near the lower boundary of the $a_\dagger$ prior.
For either profile and any \emph{fixed} scale factor $a>0$,
\begin{equation}
 x=b\left(\frac{a}{a_\dagger}-1\right)
 \xrightarrow[a_\dagger\to0^+]{}+\infty,
\end{equation}
so that $G_B(a)\to1$ and $G_B(1)\to1$. Hence
\begin{equation}
 \boxed{\lim_{a_\dagger\to0^+}R_B(a)=1,\qquad
 \lim_{a_\dagger\to0^+}\frac{\dd R_B}{\dd\ln a}=0.}
 \label{eq:earlylimit}
\end{equation}
Both families therefore approach the flat \lcdm\ background
with the same matter and radiation content and present effective
dark-energy density. The limit is pointwise for $a>0$ and is
not uniform as $a\to0$: for every finite $a_\dagger$, an earlier
negative-density phase remains. Background convergence alone
does not establish convergence of all VCDM perturbation
observables, which must be checked with the Boltzmann solver.

The background limit itself therefore imposes no strictly
positive lower cutoff on $a_\dagger$.
The point $a_\dagger=0$ represents a limiting history, since
the expression for $x$ is undefined there.
The lower sampling boundary is a prior choice and should
not itself be interpreted as an observational lower bound
on the transition parameter.

\section{Data and likelihood construction}\label{sec:data}

We constrain the cosmological models considered in this work
using CMB and BAO observations, and assess the impact of
including SN Ia distances. For this purpose, we consider
two data combinations,
\begin{align}
 \mathcal D_0&=\mathrm{CMB\!-\!SPA}+\mathrm{DESI\ DR2},\\
 \mathcal D_1&=\mathcal D_0+\mathrm{DESY5\ Dovekie}.
\end{align}
Here CMB-SPA denotes the combination of SPT, Planck, and ACT
likelihoods specified below. DESI DR2 includes the updated
Ly$\alpha$ BAO--Alcock--Paczy\'nski compression~\cite{DESI:2026lnd}, while
DESY5 Dovekie denotes the DES five-year supernova sample with
the Dovekie calibration.

For both combinations, the external optical-depth constraint
is included as a normalized Gaussian prior, rather than an
additional likelihood factor. The same calibration and
polarization-efficiency nuisance parameters are sampled
in all models.

\subsection{CMB anisotropies and lensing}

Our CMB-SPA combination follows that introduced by the SPT-3G
Collaboration~\cite{SPT-3G:2025bzu} and includes temperature
and polarization measurements from Planck, ACT, and SPT,
together with CMB lensing. We use the Planck 2018 low-$\ell$
TT likelihood and the Plik-lite high-$\ell$ TT, TE, and EE
likelihood~\cite{Planck:2019nip}, restricting the latter to
$\ell_{\max}^{TT}=1000$ and
$\ell_{\max}^{TE}=\ell_{\max}^{EE}=600$, as adopted in the
combination with ACT. We combine these data with the ACT DR6
temperature and polarization spectra~\cite{AtacamaCosmologyTelescope:2025blo}
and the SPT-3G D1 TT, TE, and EE
measurements~\cite{SPT-3G:2025bzu}.
For CMB lensing, we include the joint ACT--Planck
likelihood~\cite{Carron:2022eyg,ACT:2023dou,ACT:2023kun}
and the SPT-3G MUSE measurement of the lensing-potential
power spectrum~\cite{SPT-3G:2024atg}.

We adopt the calibration conventions of the CMB-SPA
combination, including the SPT temperature-calibration
constraint only once. Instead of a low-$\ell$ EE likelihood,
we use the normalized Gaussian prior on the reionization
optical depth $\tau$ given in Table~\ref{tab:priors}.

The likelihood combination, multipole selections, and nuisance
parameter settings are identical for all cosmological models.
Our treatment of overlap between experiments follows the
SPT-3G analysis~\cite{SPT-3G:2025bzu}, without constructing
a new joint covariance matrix. The detailed multipole
selections and nuisance settings are supplied with the
reproducibility material and are fixed analysis choices.

\subsection{DESI DR2}

We use BAO distance measurements from the second data release
of the Dark Energy Spectroscopic Instrument
(DESI DR2)~\cite{DESI:2025zgx}, including galaxies, quasars,
and the Ly$\alpha$ forest. For the Ly$\alpha$ forest,
we adopt the joint BAO and Alcock--Paczy\'nski
constraints~\cite{DESI:2026lnd}, which replace the original
BAO-only measurements. At an effective redshift $z=2.33$,
these are
\begin{align}
 D_H/r_d&=8.600\pm0.066,\\
 D_M/r_d&=39.32\pm0.33,\qquad \mathrm{corr}=0.225.
\end{align}
Here $D_H=c/H$ is the Hubble distance, $D_M$ is the transverse
comoving distance, and $r_d$ is the sound horizon at the
baryon drag epoch.

We use a Gaussian likelihood for the resulting 13 distance
measurements, retaining the galaxy and quasar BAO covariance
and including the correlation between the two Ly$\alpha$
distances. The Ly$\alpha$ constraints include the
Alcock--Paczy\'nski information extracted from the full shape
of the correlation functions, which enters our analysis
through these compressed distances. No separate galaxy
full-shape or growth-rate measurements are included.

\subsection{DESY5 with Dovekie calibration}

For \Done, we include the DESY5 Dovekie Type Ia supernova
sample, a reanalysis of the Dark Energy Survey five-year
sample incorporating the Dovekie photometric
calibration~\cite{DES:2024jxu,Popovic:2025glk,DES:2025sig}.
We use the full sample of 1820 supernovae, including the
low-redshift sample, together with its covariance matrix.

We analytically marginalize over a common additive magnitude
offset with a flat integration measure, without imposing
an external absolute-magnitude calibration. Denoting the
residual vector by $r$, the covariance matrix by $C$, and
the vector of ones by $\bm 1$, the marginalized statistic is
\begin{equation}
 Q_{\rm SN}=r^TC^{-1}r-
 \frac{(\bm1^TC^{-1}r)^2}{\bm1^TC^{-1}\bm1}
 +\ln\frac{\bm1^TC^{-1}\bm1}{2\pi}.
\end{equation}
The covariance normalization and the normalization of the
magnitude integration measure are common to all models
for a fixed supernova sample. They therefore cancel in
Bayes factors, although the absolute evidence depends
on these conventions. We compare evidences only between
models fitted to the same data combination; a change
in $\ln Z$ upon adding supernovae does not by itself
measure support for one cosmological model over another.

\section{Statistical framework}\label{sec:evidence}

\subsection{Matched priors and sampling}
\begin{table*}[t]
\caption{Normalized priors used in the analysis. The priors on the six standard cosmological parameters and four calibration parameters are identical in all eight runs. The final two entries in the right-hand column give the additional model-specific priors. $U$ denotes a uniform distribution, and $N(\mu,\sigma)$ a Gaussian distribution with mean $\mu$ and standard deviation $\sigma$. $H_0$ is expressed in $\mathrm{km\,s^{-1}\,Mpc^{-1}}$. The Gaussian prior on $\tau$ replaces the low-$\ell$ polarization likelihood.}\label{tab:priors}
\centering\small\begin{tabular}{llll}\toprule
Cosmology (all models) & Prior & Calibration / extra parameter & Prior\\\midrule
$\omega_b$ & $U(0.005,0.1)$ & $A_{\rm Planck}$ & $U(0.5,1.5)$\\
$\omega_c$ & $U(0.001,0.99)$ & $P_{\rm ACT}$ & $U(0.9,1.1)$\\
$H_0$ & $U(40,100)$ & $T_{\rm cal}$ & $N(1,0.0036)$\\
$\ln(10^{10}A_s)$ & $U(1.61,3.91)$ & $E_{\rm cal}$ & $U(0.8,1.2)$\\
$n_s$ & $U(0.8,1.2)$ & F42: $w_0$ & $U(-2,0)$\\
$\tau$ & $N(0.051,0.006)$ & Ag/Qui: $a_\dagger$ & $U(0.25,0.99)$\\\bottomrule\end{tabular}\end{table*}

The priors in Table~\ref{tab:priors} give ten sampled parameters
for \lcdm\ and eleven for each alternative. We use
MontePython~\cite{Audren:2012wb,Brinckmann:2018cvx}
with the PolyChord nested
sampler~\cite{Handley:2015fda,Handley:2015vkr}
to obtain posterior distributions and estimate the Bayesian
evidence. We adopt $200$ live points and a stopping precision
of $0.01$. The external constraints on $\tau$ and
$T_{\rm cal}$ enter as normalized Gaussian priors.
Further numerical details are given in
Appendix~\ref{app:repro}.

\subsection{Evidence and numerical support}

The Bayesian evidence averages the likelihood over the
parameter prior and therefore accounts for both the quality
of fit and the prior volume~\cite{Trotta:2008qt}.
For a model $M$ with normalized prior $\pi_M$, we define
\begin{equation}
 Z_M=\int\mathcal L_M(\theta)\pi_M(\theta)\,\dd\theta,
 \qquad \ln B_{M,\Lambda}=\ln Z_M-\ln Z_\Lambda.
 \label{eq:evidence}
\end{equation}
For equal model prior probabilities, positive
$\ln B_{M,\Lambda}$ favors $M$ over \lcdm.
The Bayes factor is a measure of relative model support,
rather than a frequentist significance. All comparisons
use the same data combination.

In our implementation, the likelihood is evaluated within
a subset $S_M$ of the adopted prior, with prior mass
$f_M=\int_{S_M}\pi_M(\theta)\,\dd\theta$.
The initialization procedure described in
Appendix~\ref{app:repro} conditions the sampling prior on
this region, yielding the evidence $Z_{M,\rm raw}$.
We also report the evidence obtained by assigning zero
likelihood outside $S_M$ under the original prior,
\begin{equation}
 Z_{M,\rm box}=f_M Z_{M,\rm raw}.
 \label{eq:volume}
\end{equation}
Here $f_M$ is estimated from the initialization counts
under the full normalized prior, including the Gaussian
priors on $\tau$ and $T_{\rm cal}$.
Table~\ref{tab:evidence} presents both normalizations.
This adjustment accounts for the excluded prior volume,
but does not recover likelihood contributions from
potentially valid points rejected because of numerical
failures. The origin of the restrictions and the estimation
of $f_M$ are discussed in Appendix~\ref{app:repro}.

The quoted evidence uncertainties include the internal
PolyChord uncertainty and, where applicable, the uncertainty
in $f_M$. They do not account for missed modes, numerical
systematics, or uncertainty in the interpretation of excluded
points. Bayes factors sharing the same \lcdm\ reference
have correlated uncertainties. A direct comparison between
two alternatives therefore uses their respective evidence
estimates, without including the uncertainty of the common
reference.

\subsection{Posterior constraints, converged fits, and AIC}

Posterior intervals are obtained from equal-tailed weighted
quantiles of the PolyChord posterior samples. Smoothed
contours are used only for visualization~\cite{Lewis:2019xzd}.

We determine best-fit parameters independently using
Metropolis--Hastings annealing to minimize
$\chi^2_{\rm phys}=-2\ln\mathcal L_{\rm phys}$, with the
likelihood normalizations specified in Sec.~\ref{sec:data}.
The external Gaussian constraints on $\tau$ and $T_{\rm cal}$
are included as priors in the Bayesian analysis and are
excluded from this data-only objective. The annealing
schedule and numerical checks are described in
Appendix~\ref{app:annealing}.

We assess numerical reproducibility through the difference
$\delta\chi^2_{\rm phys}$ between the two lowest values
obtained from completed annealing runs with distinct
starting points, requiring
\begin{equation}
 \boxed{\delta\chi^2_{\rm phys}<0.1.}\label{eq:fitcriterion}
\end{equation}
All eight model--dataset combinations satisfy this criterion.
Agreement between distinct starts does not guarantee that
the global minimum has been found. We therefore denote
the numerical best-fit estimate by
$\widehat{\chi}^2_{\rm phys,min}$.

We compare the best fits using the Akaike information
criterion (AIC), which penalizes the number of fitted
parameters~\cite{Liddle:2007fy}. For $k_M$ cosmological
and nuisance parameters, we compute
\begin{align}
 \widehat{\mathrm{AIC}}_M&=\widehat{\chi}^2_{\rm phys,min,M}+2k_M,\\
 \Delta\widehat{\mathrm{AIC}}_M&=\widehat{\mathrm{AIC}}_M
   -\widehat{\mathrm{AIC}}_{\Lambda\mathrm{CDM}}.
 \label{eq:aic}
\end{align}
We use $k_{\Lambda\mathrm{CDM}}=10$ and $k_M=11$
for each alternative, so
$\Delta\widehat{\mathrm{AIC}}_M=
\widehat{\chi}^2_{\rm phys,min,M}-
\widehat{\chi}^2_{\rm phys,min,\Lambda\mathrm{CDM}}+2$.
Negative $\Delta\widehat{\mathrm{AIC}}_M$ favors the
alternative model. Unlike the Bayesian evidence, the AIC
depends on the maximum likelihood rather than its average
over the prior. The best-fit parameters need not coincide
with the posterior maxima obtained when including the
external Gaussian priors.

\begin{figure*}[t]
\includegraphics[width=0.82\textwidth]{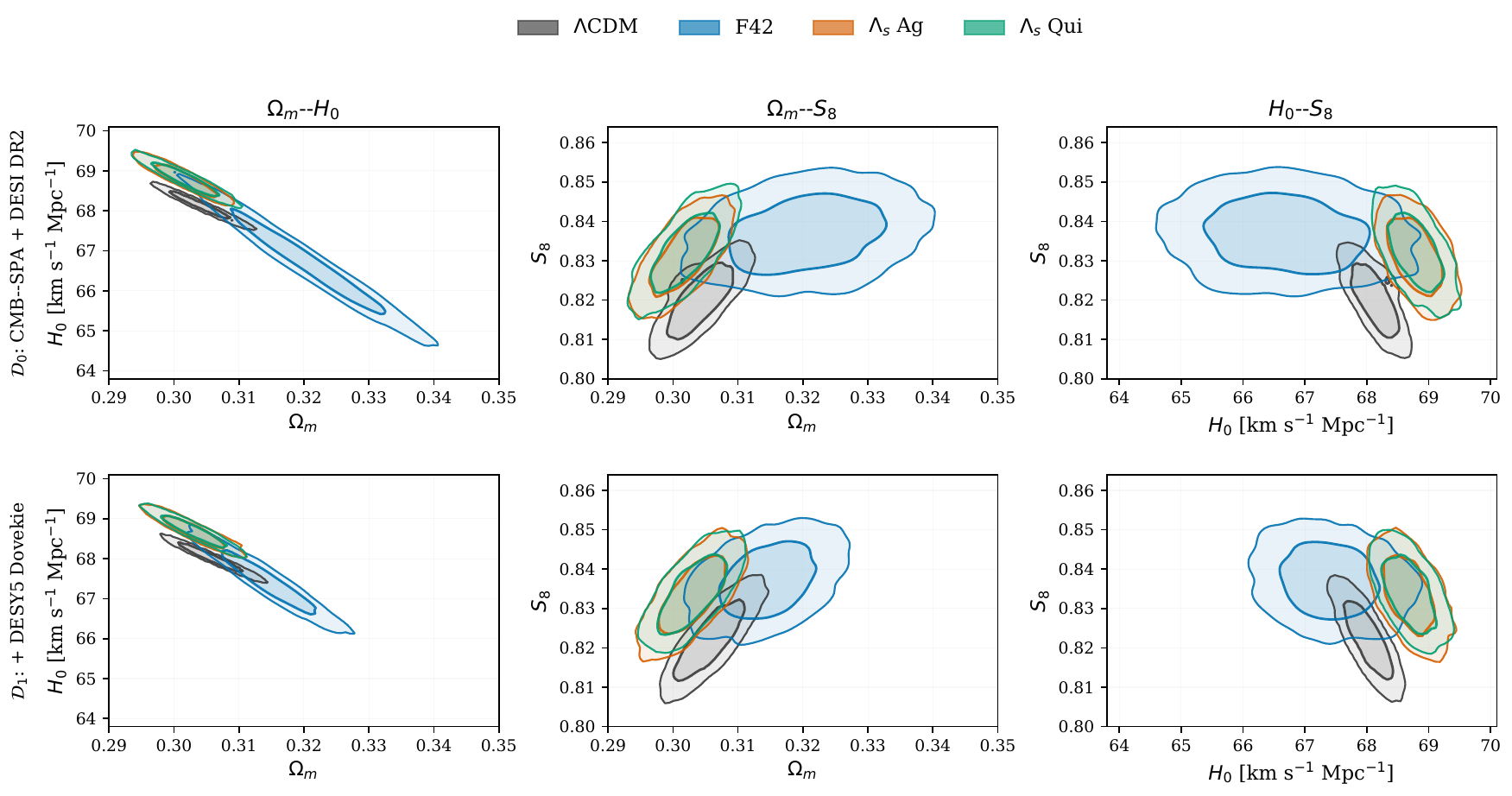}
\caption{Marginalized 68\% and 95\% credible regions
in the $\Omega_m$--$H_0$, $\Omega_m$--$S_8$, and
$H_0$--$S_8$ planes for \lcdm, F42, Ag, and Qui.
The upper panels show the constraints from $\Dzero$
and the lower panels those from $\Done$.
The axis ranges are identical in both rows.}
\label{fig:contours}
\end{figure*}

\section{Posterior constraints and model comparison}\label{sec:results}

\subsection{Cosmological parameter constraints}

\begin{table*}[t]
\caption{Marginalized constraints on the cosmological parameters, quoted as posterior medians with 68\% credible intervals. $H_0$ is in units of $\mathrm{km\,s^{-1}\,Mpc^{-1}}$, and $S_8=\sigma_8\sqrt{\Omega_m/0.3}$. The additional parameter is $w_0$ for F42 and $a_\dagger$ for Ag and Qui. The last column gives the phantom-crossing redshift $z_\times$ for F42 and the transition redshift $z_\dagger$ for Ag and Qui. For Ag, $z_\dagger$ marks the center of the transition rather than the redshift at which the density changes sign. For both Ag and Qui, we adopt the prior $a_\dagger\in[0.25,0.99]$.}
\label{tab:posteriors}\centering\small\setlength{\tabcolsep}{5pt}
\begin{tabular}{llrrrrr}\toprule
Data & Model & $H_0$ & $\Omega_m$ & $S_8$ & $w_0$ or $a_\dagger$ & $z_\times$ or $z_\dagger$\\\midrule
$\mathcal D_0$ & \lcdm & $68.13^{+0.24}_{-0.25}$ & $0.3042^{+0.0033}_{-0.0031}$ & $0.8206^{+0.0055}_{-0.0067}$ & --- & ---\\
$\mathcal D_0$ & F42 & $66.73^{+0.83}_{-0.90}$ & $0.3204^{+0.0079}_{-0.0076}$ & $0.8369^{+0.0062}_{-0.0066}$ & $-0.813^{+0.033}_{-0.029}$ & $0.51^{+0.13}_{-0.10}$\\
$\mathcal D_0$ & $\Lambda_s$ Ag & $68.78^{+0.29}_{-0.27}$ & $0.3017^{+0.0032}_{-0.0035}$ & $0.8316^{+0.0062}_{-0.0066}$ & $0.273^{+0.017}_{-0.015}$ & $2.66^{+0.22}_{-0.21}$\\
$\mathcal D_0$ & $\Lambda_s$ Qui & $68.78^{+0.27}_{-0.27}$ & $0.3020^{+0.0032}_{-0.0034}$ & $0.8324^{+0.0063}_{-0.0063}$ & $0.269^{+0.013}_{-0.012}$ & $2.72^{+0.17}_{-0.18}$\\
\midrule
$\mathcal D_1$ & \lcdm & $68.02^{+0.25}_{-0.25}$ & $0.3058^{+0.0033}_{-0.0034}$ & $0.8221^{+0.0063}_{-0.0064}$ & --- & ---\\
$\mathcal D_1$ & F42 & $67.40^{+0.53}_{-0.50}$ & $0.3144^{+0.0048}_{-0.0048}$ & $0.8367^{+0.0065}_{-0.0059}$ & $-0.838^{+0.019}_{-0.020}$ & $0.423^{+0.068}_{-0.067}$\\
$\mathcal D_1$ & $\Lambda_s$ Ag & $68.68^{+0.26}_{-0.25}$ & $0.3030^{+0.0031}_{-0.0033}$ & $0.8335^{+0.0064}_{-0.0062}$ & $0.274^{+0.016}_{-0.015}$ & $2.66^{+0.21}_{-0.20}$\\
$\mathcal D_1$ & $\Lambda_s$ Qui & $68.70^{+0.26}_{-0.27}$ & $0.3030^{+0.0034}_{-0.0032}$ & $0.8339^{+0.0062}_{-0.0064}$ & $0.268^{+0.013}_{-0.012}$ & $2.73^{+0.18}_{-0.17}$\\
\midrule\bottomrule\end{tabular}\end{table*}

Table~\ref{tab:posteriors} summarizes the marginalized
constraints, while Fig.~\ref{fig:contours} shows the
corresponding two-dimensional contours. F42 favors lower
$H_0$ and higher $\Omega_m$ than \lcdm\ for both data
combinations, with broader allowed regions in the
$\Omega_m$--$H_0$ plane. This behavior is consistent with
previous constraints on F42 from CMB, BAO, and supernova
data~\cite{Borghetto:2026ytv}, reflecting the additional
freedom in the late-time expansion history and its
degeneracy with the matter density and Hubble constant.
For \Done, we obtain
$H_0=67.40^{+0.53}_{-0.50}\,\mathrm{km\,s^{-1}\,Mpc^{-1}}$
and $\Omega_m=0.3144^{+0.0048}_{-0.0048}$ for F42,
compared with
$H_0=68.02^{+0.25}_{-0.25}\,\mathrm{km\,s^{-1}\,Mpc^{-1}}$
and $\Omega_m=0.3058^{+0.0033}_{-0.0034}$ for \lcdm.
Including supernovae substantially tightens the F42
constraints, although they remain broader than those
obtained within \lcdm.

Ag and Qui instead give nearly indistinguishable constraints
on the common cosmological parameters, favoring higher
$H_0$ and lower $\Omega_m$ than \lcdm.
For \Done, Ag gives
$H_0=68.68^{+0.26}_{-0.25}\,\mathrm{km\,s^{-1}\,Mpc^{-1}}$
and $\Omega_m=0.3030^{+0.0031}_{-0.0033}$, while Qui gives
$H_0=68.70^{+0.26}_{-0.27}\,\mathrm{km\,s^{-1}\,Mpc^{-1}}$
and $\Omega_m=0.3030^{+0.0034}_{-0.0032}$.
Their similar constraints are consistent with the fact
that both profiles approach the same constant-density
limits away from the transition and differ mainly through
the localized enhancement in Ag. Moreover, both favor
$a_\dagger\simeq0.27$, placing this difference within a
narrow interval around $z_\dagger\simeq2.7$.

The distinction from \lcdm\ is more apparent in the joint
constraints than in the individual parameter intervals.
In the $\Omega_m$--$H_0$ plane, the Ag and Qui contours
have a similar orientation to those of \lcdm, but are
displaced towards higher $H_0$ and lower $\Omega_m$.
All three alternatives also favor higher $S_8$: for
\Done, the central values are approximately $0.837$ for
F42 and $0.834$ for Ag and Qui, compared with $0.822$
for \lcdm. These correlated shifts illustrate how the
assumed dark-energy history affects the inferred expansion
and growth parameters.

Including the DESY5 Dovekie supernova sample affects F42
more strongly than Ag and Qui. To quantify these changes,
we report in Table~\ref{tab:shifts} the shifts in the
posterior means,
\[
\delta_\theta=
\frac{\overline\theta_1-\overline\theta_0}{\sigma_0(\theta)},
\]
where $\overline\theta_0$ and $\overline\theta_1$ are the
posterior means obtained with \Dzero\ and \Done,
respectively, and $\sigma_0(\theta)$ is the marginal
standard deviation for \Dzero. For F42, the mean $H_0$
increases by approximately
$0.70\,\mathrm{km\,s^{-1}\,Mpc^{-1}}$, while $\Omega_m$
and $w_0$ decrease by $0.0060$ and $0.026$, respectively.
All reported shifts are smaller than one standard
deviation of the \Dzero\ posterior. The transition
parameter in Ag and Qui changes by less than $0.1$
times its standard deviation under \Dzero.

Since \Done\ includes \Dzero, these shifts describe
the effect of adding supernovae and should not be
interpreted as tension significances between independent
datasets.

\begin{figure*}[t]
\includegraphics[width=\textwidth]{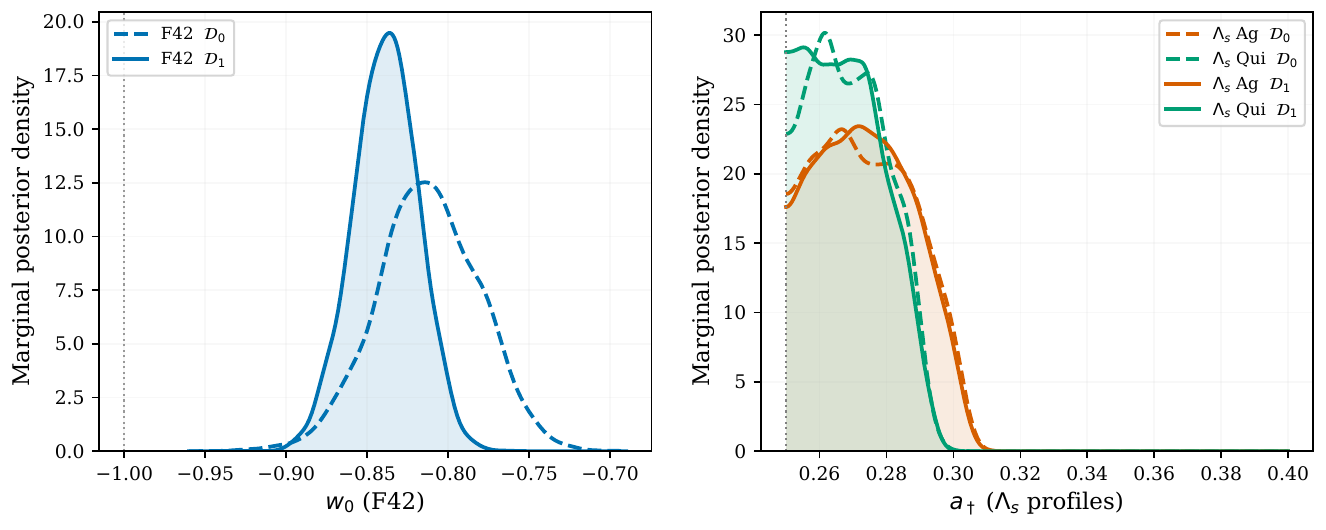}
\caption{One-dimensional posterior distributions of
$w_0$ for F42 (left) and $a_\dagger$ for Ag and Qui
(right). Dashed curves show the constraints from
$\Dzero$, while solid curves with shading show those
from $\Done$. The vertical line in the right panel
marks the lower prior boundary, $a_\dagger=0.25$.}
\label{fig:parameters}
\end{figure*}

\begin{table}[t]
\caption{Shifts in the posterior means between \Dzero\ and \Done, expressed in units of the marginal standard deviation for \Dzero. Entries give $(\overline\theta_1-\overline\theta_0)/\sigma_0(\theta)$. Since \Done\ includes \Dzero, these shifts should not be interpreted as tension significances. The last column refers to $w_0$ for F42 and $a_\dagger$ for Ag and Qui.}
\label{tab:shifts}\centering\small\setlength{\tabcolsep}{4.7pt}
\begin{tabular}{lrrrr}\toprule
Model & $\Omega_m$ & $H_0$ & $S_8$ & Model parameter \\ \midrule
$\Lambda$CDM & +0.48 & -0.46 & +0.30 & --- \\
F42 & -0.79 & +0.83 & +0.01 & -0.84 \\
$\Lambda_s$ Ag & +0.42 & -0.37 & +0.34 & -0.01 \\
$\Lambda_s$ Qui & +0.34 & -0.34 & +0.22 & -0.08 \\
\bottomrule\end{tabular}\end{table}

Figure~\ref{fig:parameters} shows that including the
DESY5 Dovekie supernova sample also tightens the F42
constraint, shifting $w_0$ from
$-0.813^{+0.033}_{-0.029}$ to
$-0.838^{+0.019}_{-0.020}$. Using
Eq.~\eqref{eq:crossing}, the corresponding
phantom-crossing redshifts are
\begin{align}
 z_\times(\Dzero)&=0.51^{+0.13}_{-0.10},\\
 z_\times(\Done)&=0.423^{+0.068}_{-0.067}.
\end{align}
Within F42, including supernovae therefore favors a
later phantom crossing and reduces its uncertainty.
The Ag and Qui transition constraints remain nearly
unchanged and lie close to the lower boundary of the
adopted prior, $a_\dagger=0.25$.

\begin{figure*}[t]
\includegraphics[width=\textwidth]{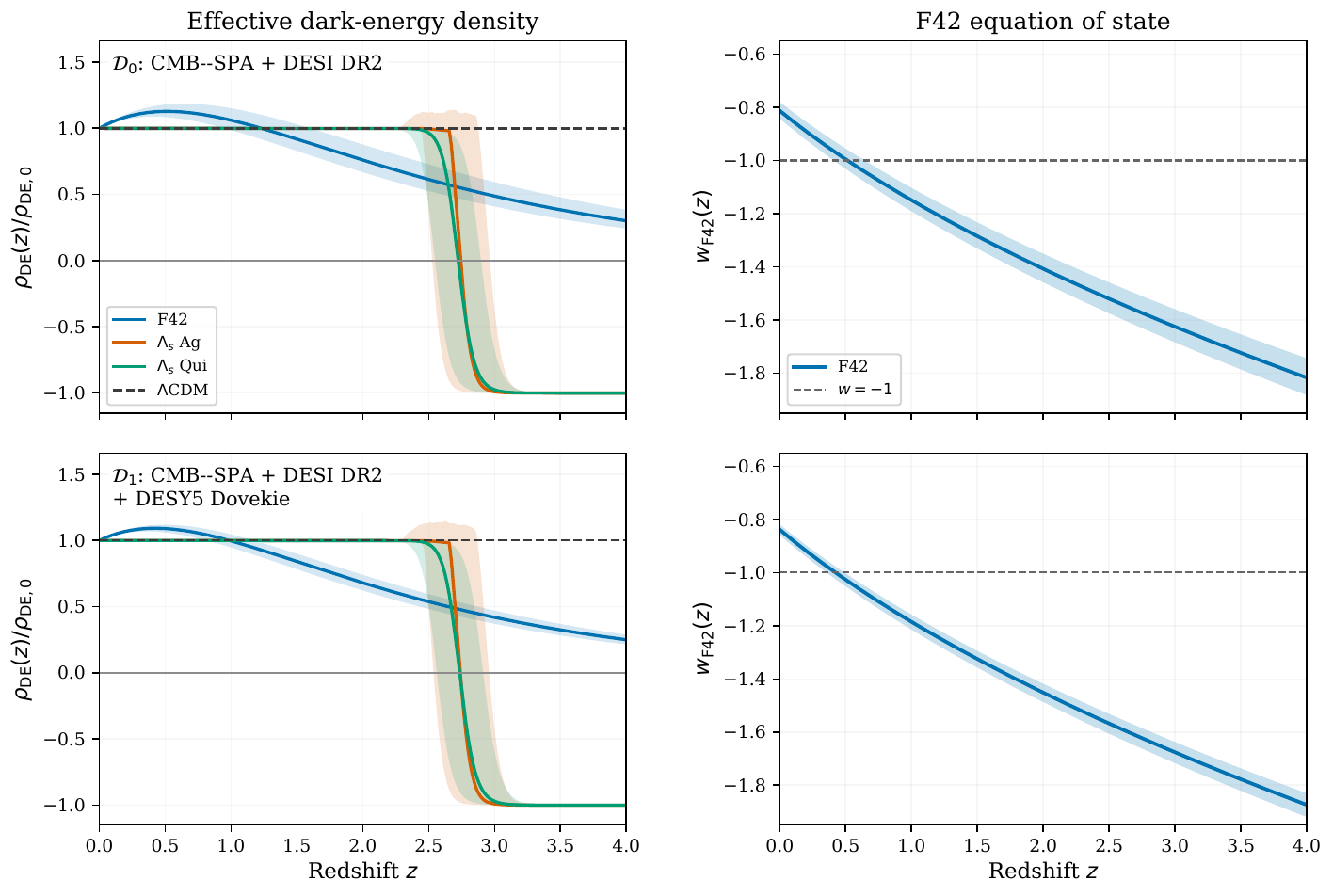}
\caption{Evolution of the effective dark-energy density
normalized to its present value (left) and the F42
equation of state (right). The upper panels show
$\Dzero$ and the lower panels $\Done$. Curves and
shaded bands give the median and central 68\% credible
interval at each redshift. The Ag and Qui constraints
assume $a_\dagger\in[0.25,0.99]$.}
\label{fig:histories}
\end{figure*}

Figure~\ref{fig:histories} shows the corresponding
dark-energy evolution. Ag has a localized density
enhancement around the transition, while Qui evolves
monotonically from negative to positive density.
F42 maintains a positive density that becomes negligible
at early times for $w_0<0$, with an equation of state
that crosses $w=-1$. Thus, the similar constraints
on the common cosmological parameters in Ag and Qui
do not imply identical dark-energy histories.

Complementary Cobaya MCMC analyses for Qui give
broadly consistent posterior constraints, as reported
in Appendix~\ref{app:mcmc}. These analyses use different
DESI Ly$\alpha$, low-$\ell$ polarization,
and nonlinear settings, and are therefore kept separate
from the matched model comparison.

\subsection{Model comparison}
\label{sec:matchedfit}

Table~\ref{tab:aic} presents the best-fit estimates for all
four models, obtained with the same annealing procedure
and satisfying the numerical reproducibility criterion
in Eq.~\eqref{eq:fitcriterion}. All three alternatives
give a lower $\chi^2_{\rm phys}$ than \lcdm.
The AIC assesses whether this improvement compensates
for the additional fitted parameter.

For \Dzero, the three $\Delta\widehat{\mathrm{AIC}}$
values are close to zero, indicating no clear preference
over \lcdm\ according to this criterion. For \Done,
we obtain $-4.02$ for F42, $-0.25$ for Ag, and $-0.24$
for Qui. F42 therefore has the lowest AIC among the
models considered when supernovae are included,
whereas Ag and Qui remain nearly indistinguishable
from \lcdm\ on this criterion. The difference between
Ag and Qui is also smaller than the tolerance adopted
in our numerical reproducibility check.

We next examine the Bayesian evidence, which averages
the likelihood over the adopted priors rather than
using its maximum value. In particular, the evidence
includes the external Gaussian priors on $\tau$ and
$T_{\rm cal}$, whereas these constraints are excluded
from the likelihood used to compute the AIC.

\begin{table*}[t]
\caption{Best-fit $\chi^2_{\rm phys}$ and AIC differences
relative to \lcdm\ for the two data combinations.
The best fits are obtained using the same annealing
procedure and satisfy Eq.~\eqref{eq:fitcriterion}.
The external Gaussian priors on $\tau$ and $T_{\rm cal}$
are excluded from the optimized likelihood.
Here $k$ is the number of fitted cosmological and nuisance
parameters. Negative $\Delta\widehat{\mathrm{AIC}}$
favors the alternative model.}
\label{tab:aic}
\centering\small
\setlength{\tabcolsep}{8pt}
\begin{tabular}{lrrrrr}\toprule
 & & \multicolumn{2}{c}{$\Dzero$}
     & \multicolumn{2}{c}{$\Done$} \\
\cmidrule(lr){3-4}\cmidrule(lr){5-6}
Model & $k$ & $\widehat{\chi}^2_{\rm phys,min}$ & $\Delta\widehat{\mathrm{AIC}}$
            & $\widehat{\chi}^2_{\rm phys,min}$ & $\Delta\widehat{\mathrm{AIC}}$\\\midrule
$\Lambda$CDM & 10 & 615.343 & --- & 2259.501 & --- \\
F42 & 11 & 612.942 & $-0.40$ & 2253.484 & $-4.02$ \\
Ag  & 11 & 613.255 & $-0.09$ & 2257.248 & $-0.25$ \\
Qui & 11 & 613.295 & $-0.05$ & 2257.265 & $-0.24$ \\
\bottomrule
\end{tabular}
\end{table*}

\begin{table*}[t]
\caption{Bayesian evidence estimates and log Bayes factors
relative to \lcdm\ for each data combination.
Native estimates use the prior restricted and normalized
to the region where the likelihood is evaluated.
The ``box'' estimates assign zero likelihood outside
this region, as defined in Eq.~\eqref{eq:volume}.
Uncertainties include the internal PolyChord error and,
for the ``box'' estimates, the uncertainty in the
fraction of prior mass retained. For Ag and Qui,
we adopt $a_\dagger\in[0.25,0.99]$.}\label{tab:evidence}
\centering\small\setlength{\tabcolsep}{9pt}
\begin{tabular}{lrrrr}\toprule
Model & $\ln Z_{\rm raw}$ & $\widehat f$ & $\Delta\ln Z_{\rm raw}$ & $\Delta\ln Z_{\rm box}$\\\midrule
\multicolumn{5}{c}{$\mathcal D_0$: CMB-SPA + DESI DR2}\\\midrule
\lcdm & $-351.78\pm0.42$ & 0.0991 & --- & ---\\
F42 & $-351.18\pm0.44$ & 0.1028 & $0.60\pm0.61$ & $0.63\pm0.61$\\
Ag & $-349.59\pm0.43$ & 0.0717 & $2.19\pm0.60$ & $1.87\pm0.61$\\
Qui & $-350.41\pm0.43$ & 0.0665 & $1.37\pm0.60$ & $0.97\pm0.61$\\\midrule
\multicolumn{5}{c}{$\mathcal D_1$: $\mathcal D_0$ + DESY5 Dovekie}\\\midrule
\lcdm & $-1173.91\pm0.42$ & 0.1001 & --- & ---\\
F42 & $-1171.70\pm0.44$ & 0.0981 & $2.21\pm0.61$ & $2.19\pm0.61$\\
Ag & $-1172.36\pm0.43$ & 0.0691 & $1.54\pm0.61$ & $1.17\pm0.61$\\
Qui & $-1171.62\pm0.43$ & 0.0613 & $2.28\pm0.60$ & $1.79\pm0.61$\\
\bottomrule\end{tabular}\end{table*}

\begin{figure*}[t]
\includegraphics[width=\textwidth]{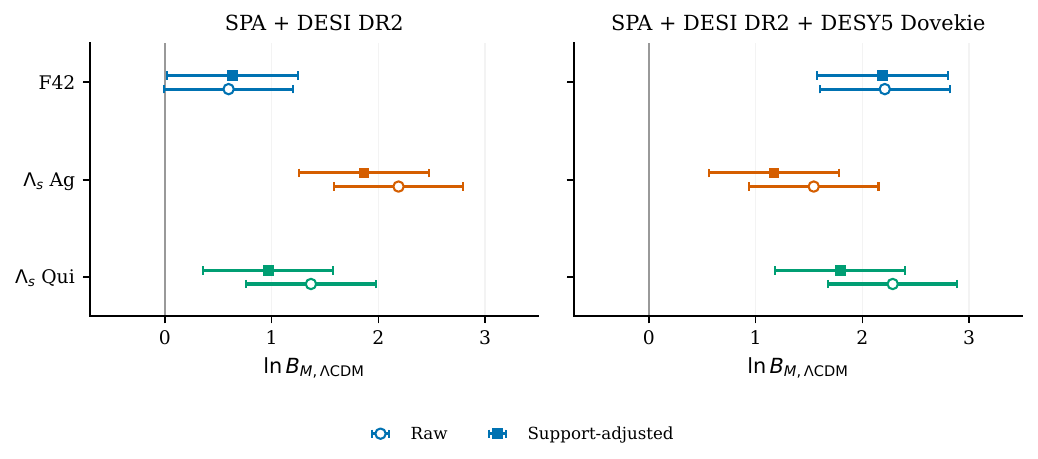}
\caption{Log Bayes factors relative to \lcdm\ for
$\Dzero$ (left) and $\Done$ (right). Open circles
show the native estimates, and squares show the
``box'' estimates defined in Eq.~\eqref{eq:volume}.
Error bars include the internal PolyChord uncertainty
and, for the ``box'' estimates, the uncertainty in
the retained prior mass. Numerical and likelihood
systematics are not included. All estimates use
the priors in Table~\ref{tab:priors}, with
$a_\dagger\in[0.25,0.99]$ for Ag and Qui.} 
\label{fig:evidence}
\end{figure*}

Table~\ref{tab:evidence} and Fig.~\ref{fig:evidence} show
the log Bayes factors relative to \lcdm\ for each data
combination. We report both the native and ``box''
estimates defined in Eq.~\eqref{eq:volume}, which differ
in their treatment of the prior volume outside the
evaluated likelihood region.

For \Dzero, Ag has the largest evidence under both
conventions. The ``box'' log Bayes factors are
$\BFAgZero$ for Ag, $\BFFZero$ for F42, and
$\BFQuiZero$ for Qui. The central values favor the
alternatives over \lcdm, with the strongest preference
for Ag, although the differences between the
alternatives remain modest.

For \Done, the native log Bayes factors are
$\RAWFOne$ for F42, $\RAWAgOne$ for Ag, and
$\RAWQuiOne$ for Qui. The corresponding ``box''
values are $\BFFOne$, $\BFAgOne$, and $\BFQuiOne$.
F42 has the largest central evidence under the
``box'' convention, while Qui has a slightly larger
native evidence. The direct difference
$\ln Z_{\rm F42}-\ln Z_{\rm Qui}$ is
$-0.07\pm0.61$ for the native estimates and
$0.40\pm0.62$ for the ``box'' estimates.
Neither comparison distinguishes F42 from Qui
within the quoted numerical uncertainties.

The change in normalization affects Ag and Qui more
strongly because the evaluated region contains about
$6$--$7\%$ of their prior mass, compared with approximately
$10\%$ for \lcdm\ and F42. This changes the evidence
but leaves the best-fit likelihood unchanged.
Comparing \Dzero\ with \Done, the central ``box''
log Bayes factors increase by $1.56$ for F42 and
$0.82$ for Qui, and decrease by $0.69$ for Ag.
Thus, including the DESY5 Dovekie supernova sample
strengthens the relative support for F42 and Qui
and weakens it for Ag at the level of these central
estimates. These changes refer to Bayes factors
relative to \lcdm\ within each data combination,
rather than differences between absolute evidences
for different datasets.

Taken together, the AIC and Bayesian evidence show
that an improved fit does not uniquely select a
dark-energy history. F42 gives the lowest AIC for
\Done, but its evidence is not clearly distinguished
from that of Qui. The evidence ranking also depends
on the adopted treatment of the excluded prior volume.

\section{Discussion: what would discriminate the models?}
\label{sec:limits}

The models describe different physical
histories. F42 has a continuously evolving positive
dark-energy density with a possible phantom crossing,
whereas Ag and Qui describe smooth transitions from
negative to positive effective density. Both sign-changing
profiles approach the \lcdm\ background in the
early-transition limit. However, the posterior obtained
with $a_\dagger\geq0.25$ does not constrain the excluded
region or establish a prior-independent lower bound on
$a_\dagger$. Runs exploring $0.10<a_\dagger<0.25$
are treated separately and do not enter the evidence
comparison presented here.

Within VCDM, the background and linear perturbations
follow from the same gravitational theory, with only
two propagating gravitational degrees of
freedom~\cite{DeFelice:2020eju,DeFelice:2022uxv}.
This allows the proposed histories to be tested through
structure growth and lensing as well as expansion
distances. The similar constraints on the common
cosmological parameters, particularly for Ag and Qui,
motivate additional measurements of structure formation
and its redshift evolution.

Euclid combines weak gravitational lensing and
spectroscopic galaxy clustering, including BAO and
redshift-space distortions, to probe geometry, growth,
and gravitational potentials~\cite{Euclid:2024yrr}.
These complementary measurements may help distinguish
histories that provide similar fits to current distance
data. Such an analysis requires sufficiently accurate
predictions for VCDM perturbations and nonlinear
structure formation. We use the same HMcode-2020
prescription~\cite{Mead:2020vgs} for all models, but
its accuracy for these VCDM histories has not been
independently established. Likewise, applying the
compressed DESI distances assumes that the distance
extraction remains applicable to the alternative
cosmologies.

The evidence estimates also depend on the treatment
of the likelihood evaluation domain. The Commander
and MUSE implementations impose restrictions, and
the native and ``box'' estimates correspond to the
prior conventions defined in Sec.~\ref{sec:evidence}.
Assigning zero likelihood outside the evaluated
region does not recover any missing likelihood
contributions. The associated checks and statistical
uncertainties are described in Appendix~\ref{app:repro}.
These limitations should be considered when interpreting
the relative support for the models.

\section{Conclusions}

We have compared three histories of cosmic acceleration---F42,
Ag, and Qui---with \lcdm, each introducing one additional
cosmological parameter. All models are implemented within
VCDM, which determines the background and linear perturbations
without introducing an additional propagating gravitational
degree of freedom.

Using CMB-SPA and DESI DR2 distances, including the joint
Ly$\alpha$ BAO--Alcock--Paczy\'nski constraints, we find that
all three alternatives improve the best-fit $\chi^2_{\rm phys}$
relative to \lcdm, both with and without the DESY5 Dovekie
supernova sample. For \Done, the estimated
$\Delta\widehat{\mathrm{AIC}}$ values are $-4.02$ for F42,
$-0.25$ for Ag, and $-0.24$ for Qui. F42 therefore has
the lowest AIC and also the largest central Bayesian
evidence under the ``box'' convention. However, the
evidence differences do not establish a robust preference
among the three alternatives within the quoted numerical
uncertainties and adopted prior conventions.

The models lead to different cosmological constraints.
F42 favors lower $H_0$ and higher $\Omega_m$ than \lcdm,
with broader uncertainties, while Ag and Qui give nearly
identical constraints, shifting $H_0$ upwards and $\Omega_m$
downwards. All three alternatives favor higher $S_8$.
Including supernovae tightens the F42 constraints and
favors a later phantom crossing, while the Ag and Qui
transition constraints change little.

For Ag and Qui, the early-transition limit approaches
the \lcdm\ background. The imposed lower prior boundary
$a_\dagger=0.25$ is therefore not an observational lower
bound. Our constraints and evidence estimates apply to
the adopted prior range; extending it would require
evaluating the likelihood in the additional region.

Future measurements combining expansion distances,
galaxy clustering, and weak gravitational lensing,
particularly from Euclid~\cite{Euclid:2024yrr}, may
help distinguish these histories through their effects
on expansion and structure growth. This requires
accurate predictions for both linear perturbations
and nonlinear structure formation within VCDM.

\begin{acknowledgments}
EDV is supported by a Royal Society Dorothy Hodgkin Research Fellowship.
The work of J.Z.\ was supported by the Japanese Government (MEXT) Scholarship Program. LAE\ acknowledges support from T\"{U}B\.{I}TAK through a postdoctoral researcher fellowship associated with Grant No.~124N627. 
The authors acknowledge the use of High-Performance Computing resources from the IT Services at the University of Sheffield.
The numerical calculations presented in this work were performed in part using the Heian computing system at the Yukawa Institute for Theoretical Physics, Kyoto University.
\end{acknowledgments}

\appendix

\section{Evidence normalization and numerical provenance}\label{app:repro}

\subsection{Admitted prior measure}

Let $S_M$ denote the region accepted by the numerical
implementation and
$f_M=\int_{S_M}\pi_M(\theta)\,\dd\theta$ its prior mass.
The initialization used in these runs assigns unit prior
mass to this region, corresponding to the conditional
prior $\pi_{M,S}=\pi_M\mathbf{1}_{S_M}/f_M$, where
$\mathbf{1}_{S_M}$ is the indicator function of $S_M$.
The native evidence is therefore
$Z_{M,\rm raw}=f_M^{-1}\int_{S_M}\mathcal L_M\pi_M\,\dd\theta$.
If the likelihood is defined over the full adopted prior,
its evidence satisfies
\begin{equation}
 Z_{M,\rm declared}=f_M Z_{M,\rm raw}+
 \int_{S_M^c}\mathcal L_M(\theta)\pi_M(\theta)\,\dd\theta.
 \label{eq:missing-support}
\end{equation}
Equation~\eqref{eq:volume} follows by assigning zero
likelihood outside $S_M$. It does not estimate the
likelihood contribution from the excluded region.
In particular, failure of a numerical calculation
does not establish that the corresponding parameters
are physically excluded.

We estimate $f_M$ from the fraction of accepted
initialization attempts. In the implementation used
here, the counter \texttt{ndiscarded} records the
total number of attempts, $N_{\rm att}$, including
the $N_{\rm acc}$ accepted points. Assuming independent
draws from the prior, the estimated fraction and its
approximate logarithmic uncertainty are
\begin{equation}
 \widehat f_M=\frac{N_{\rm acc}}{N_{\rm att}},\qquad
 \sigma_{\ln\widehat f}\simeq
 \sqrt{\frac{1-\widehat f_M}{N_{\rm acc}}}.
 \label{eq:f-est}
\end{equation}
The target was 400 accepted points; parallel execution
resulted in 404--409 accepted points per run.
The fraction $\widehat f_M$ is therefore a statistical
estimate of the accepted prior mass.

To examine the reasons for rejection, we evaluated
40 independent points drawn from the prior for each
run, giving 320 points in total. Of these, 274 were
excluded by the evaluation-domain restrictions of
the Commander low-$\ell$ TT or MUSE likelihoods,
11 had no real-valued past background solution,
eight encountered unresolved numerical failures,
and 27 were accepted. This check identifies the
dominant sources of rejection, but does not bound
the missing contribution in Eq.~\eqref{eq:missing-support}.

\subsection{Implementation and statistical uncertainty}

The runs used graded slice repeats $[30,30]$, a compression
factor of $0.8$, and clustering disabled. CMB spectra were
computed up to $\ell_{\max}=9000$. The Gaussian priors on
$\tau$ and $T_{\rm cal}$ were implemented through inverse
cumulative distribution functions mapping uniform samples
from the unit cube to the corresponding normalized
Gaussian distributions.

The BBN interpolation routine extrapolates outside the
range covered by its table. Although the posterior samples
lie within the standard baryon-density range, the treatment
of the broader prior remains relevant to the evaluated
likelihood region and the evidence.

As a numerical cross-check, we re-evaluated 32 parameter
points from the posterior samples, covering all eight runs,
using the same CLASS implementation. The resulting total
log-likelihoods agreed with the stored values to within
$2.61\times10^{-3}$. This checks reproducibility at the
selected points but does not replace independent
nested-sampling runs.

We analyze one nested-sampling run for each model and
data combination. The internal PolyChord uncertainties
do not account for missed modes, incomplete mixing in
the slice sampling, or sensitivity to the random seed.
The weight diagnostic in Table~\ref{tab:runs} measures
the concentration of sample weights rather than the
number of statistically independent samples.
Assuming independent runs, their evidence uncertainties
are combined in quadrature. Bayes factors sharing the
same \lcdm\ reference have correlated uncertainties.
For the ``box'' estimates, we also include the uncertainty
in the accepted prior fraction from
Eq.~\eqref{eq:f-est}.

\begin{table}[!tb]
\caption{Nested-sampling diagnostics for the eight runs.
The initialization target was $n_{\rm prior}=400$ accepted
points; parallel execution can yield slightly more.
$N_{\rm att}$ is the total number of initialization attempts,
recorded by \texttt{ndiscarded} in the PolyChord implementation
used here. $N_{\rm w}=1/\sum_i\widetilde w_i^2$, where
$\widetilde w_i$ are the normalized sample weights, measures
weight concentration rather than the number of statistically
independent samples. All final statistics files report zero
live points and zero active clusters.}
\label{tab:runs}\centering\small
\begin{tabular}{llrrr}\toprule Model & Data & Dead & $N_{\rm acc}/N_{\rm att}$ & $N_{\rm w}$\\\midrule
\lcdm & $\mathcal D_0$ & 10682 & 405/4087 & 1726\\
\lcdm & $\mathcal D_1$ & 10624 & 405/4045 & 1753\\
F42 & $\mathcal D_0$ & 11081 & 406/3949 & 1612\\
F42 & $\mathcal D_1$ & 11124 & 407/4148 & 1601\\
$\Lambda_s$ Ag & $\mathcal D_0$ & 10730 & 409/5702 & 1471\\
$\Lambda_s$ Ag & $\mathcal D_1$ & 10935 & 405/5865 & 1584\\
$\Lambda_s$ Qui & $\mathcal D_0$ & 10968 & 404/6072 & 1589\\
$\Lambda_s$ Qui & $\mathcal D_1$ & 10838 & 404/6590 & 1559\\\bottomrule\end{tabular}\end{table}

The numerical record contains weighted sample summaries,
checks of the analysis settings, inputs for the tables
and figures, and the best-fit parameters and diagnostics
from the Metropolis--Hastings annealing runs.
The full chains and executables are retained separately.

Posterior contours are obtained from weighted
two-dimensional kernel density estimates. For the
one-dimensional $a_\dagger$ distributions, we use
reflection at the lower prior boundary to avoid smoothing
probability outside the allowed range. Credible intervals
are calculated directly from the weighted samples.
The figures can therefore be regenerated without further
Boltzmann calculations.

\section{Annealing convergence and likelihood convention}
\label{app:annealing}

The best-fit search used the same three-stage annealing
sequence, with inverse temperatures $\beta=1,200,1000$,
for all eight model--dataset combinations. The proposal
covariance was estimated from the posterior samples
for each combination. Final-stage candidates were
re-evaluated using the unscaled physical-data likelihood,
with no change detected in the recorded checks.
At each selected best-fit point, the sum of the
experiment-level contributions to $-2\ln\mathcal L_{\rm phys}$
reproduces $\widehat{\chi}^2_{\rm phys,min}$.

For all eight combinations, the two lowest values
obtained from completed annealing runs with distinct
starting points differ by less than $0.1$ in
$\chi^2_{\rm phys}$, satisfying
Eq.~\eqref{eq:fitcriterion}. The largest difference
is $0.0753$. Results from less successful starting
points are retained in the numerical record.
This agreement demonstrates reproducibility between
the selected runs, but does not exclude a lower
minimum elsewhere in parameter space. Additional
optimizations at fixed transition parameters are
not used in the model comparison.

The Gaussian constraints on $\tau$ and $T_{\rm cal}$
are excluded from the likelihood used to compute
the AIC. The Gaussian penalties relative to the
prior peaks can be substantial at the physical-data
best fits, so these points need not coincide with
the posterior maxima of the Bayesian analysis.
The numerical record contains the full-precision
best-fit parameters and separate likelihood and
Gaussian-prior contributions.

\begin{table*}[t]
\caption{Posterior constraints for Qui from the main
PolyChord analysis and the complementary Cobaya MCMC
robustness tests. Values are posterior medians with
equal-tailed 68\% credible intervals. The Cobaya runs
use the standard DESI DR2 BAO measurements, replace
the Gaussian $\tau$ prior with the low-$\ell$ SRoll2
EE likelihood, and use Halofit instead of HMcode-2020.
Their convergence statistics are $R-1=0.0090$ without
supernovae and $0.0094$ with DESY5 Dovekie.
These runs do not enter the model comparison.}
\label{tab:luiscomp}\centering\small\setlength{\tabcolsep}{5pt}
\begin{tabular}{llrrrr}\toprule
SNe & Analysis & $H_0$ & $\Omega_m$ & $S_8$ & $a_\dagger$\\\midrule
None & PolyChord (Qui) & $68.78^{+0.27}_{-0.27}$ & $0.3020^{+0.0032}_{-0.0034}$ & $0.8324^{+0.0063}_{-0.0063}$ & $0.269^{+0.013}_{-0.012}$\\
None & Cobaya MCMC (Qui) & $68.93^{+0.28}_{-0.28}$ & $0.3000^{+0.0035}_{-0.0033}$ & $0.8310^{+0.0067}_{-0.0065}$ & $0.270^{+0.016}_{-0.013}$\\
\midrule
DESY5 Dovekie & PolyChord (Qui) & $68.70^{+0.26}_{-0.27}$ & $0.3030^{+0.0034}_{-0.0032}$ & $0.8339^{+0.0062}_{-0.0064}$ & $0.268^{+0.013}_{-0.012}$\\
DESY5 Dovekie & Cobaya MCMC (Qui) & $68.80^{+0.29}_{-0.28}$ & $0.3016^{+0.0034}_{-0.0034}$ & $0.8332^{+0.0064}_{-0.0065}$ & $0.269^{+0.016}_{-0.013}$\\
\bottomrule\end{tabular}
\end{table*}

\section{Transition-prior extensions}\label{app:regions}

For a fixed transition scale, we define the marginal
likelihood over the remaining parameters as
\begin{equation}
 m(a_\dagger)=\int\mathcal L(\vartheta,a_\dagger)
                  \pi(\vartheta)\,\dd\vartheta,
\end{equation}
where $\pi(\vartheta)$ is normalized and independent
of $a_\dagger$. For a uniform transition prior over
an interval $I$ of width $|I|$, the evidence is
$Z_I=|I|^{-1}\int_I m(a_\dagger)\,\dd a_\dagger$.
The evidence over a union of disjoint intervals is
therefore the average of their evidences weighted
by their prior probabilities, provided the likelihood,
remaining priors, and treatment of excluded regions
are unchanged.

The main analysis uses $A=[0.25,0.99]$. Extending
the prior to include $B=[0.10,0.25)$ gives
\begin{equation}
 Z_{[0.10,0.99]}=\frac{0.74}{0.89}Z_A+
                       \frac{0.15}{0.89}Z_B.
 \label{eq:regions}
\end{equation}
Each $Z_I$ is calculated with the normalized prior
on that interval. These expressions cannot be applied
directly to native evidences conditioned on different
accepted regions without accounting for their prior
masses. Since $Z_B\geq0$,
\begin{equation}
 \ln Z_{[0.10,0.99]}-\ln Z_A
 \geq\ln(0.74/0.89)=-0.18457\ldots.
\end{equation}
This extension can therefore reduce the log evidence
by at most approximately $0.185$, although it may
increase it. The bound does not determine the
unknown evidence $Z_B$.

Extending the uniform prior to $0<a_\dagger<0.99$
requires the evidence over $C=(0,0.25)$:
\begin{equation}
 Z_{(0,0.99)}=\frac{0.74}{0.99}Z_A+
                    \frac{0.25}{0.99}Z_C.
 \label{eq:zeroprior}
\end{equation}
The maximum possible decrease in log evidence is then
$-\ln(0.74/0.99)=0.29105\ldots$.
Results for $[0.10,0.25)$ alone do not determine
$Z_C$, since the contribution from $(0,0.10)$
must also be included. Neither Eq.~\eqref{eq:regions}
nor Eq.~\eqref{eq:zeroprior} has been applied to the
reported evidence estimates.

Approximating the evidence for sufficiently early
transitions by $Z_\Lambda$ requires agreement with
the \lcdm\ likelihood over the parameter region
contributing to the evidence, with compatible priors
and likelihood support. Agreement at a single point,
or between optimized likelihood values, is insufficient.
A logarithmic prior proportional to $1/a_\dagger$
is not normalizable on $(0,0.99)$ and therefore
requires a finite lower cutoff or a different proper
prior for evidence calculations.

For the extension in Eq.~\eqref{eq:regions}, the
posterior probabilities of the two regions are
\begin{align}
 P(A\mid D)&=\frac{0.74Z_A}{0.74Z_A+0.15Z_B},\\
 P(B\mid D)&=1-P(A\mid D).
\end{align}
When combining separately normalized regional samples,
their weights must be multiplied by these probabilities.
Equal weighting would generally be incorrect.
Since the additional regional evidences have not
been included, we do not report credible intervals
or observational bounds for the extended priors.

\section{Complementary Cobaya MCMC posterior cross-check}\label{app:mcmc}

We test the robustness of our results for a subset of
the cases considered in the main analysis, focusing
on Qui with and without the DESY5 Dovekie supernova
sample. We perform two MCMC analyses using
Cobaya~\cite{Torrado:2020dgo}, each with four chains.

Both analyses combine SPT, Planck, and ACT observations
with the standard DESI DR2 BAO
measurements~\cite{DESI:2025zgx}, without replacing
the original Ly$\alpha$ distances by the joint
BAO--AP constraints. We replace the external Gaussian
prior on $\tau$ with the Planck low-$\ell$ SRoll2 EE
likelihood~\cite{Pagano:2019tci}, and replace
HMcode-2020~\cite{Mead:2020vgs} with
Halofit~\cite{Smith:2002dz,Takahashi:2012em} for the
nonlinear corrections. As in the main analysis,
we assume three degenerate massive neutrino species
with a fixed total mass $\sum m_\nu=0.06\,\mathrm{eV}$.
The convergence statistics are $R-1=0.0090$ and
$0.0094$ without and with the DESY5 Dovekie supernova
sample, respectively. We discard the initial 30\%
of cumulative sample weight from each chain as burn-in.

We obtain $a_\dagger=0.270^{+0.016}_{-0.013}$ without
supernovae and $0.269^{+0.016}_{-0.013}$ with DESY5
Dovekie, broadly consistent with the PolyChord
constraints. The inferred transition scale is therefore
stable under the combined changes tested here.
Both analyses retain the lower prior boundary
$a_\dagger=0.25$, so this agreement does not establish
an observational lower bound on the transition parameter.
These additional runs are used to assess the robustness
of the posterior constraints and do not enter the
model comparison.

\bibliographystyle{apsrev4-1}
\bibliography{references}
\end{document}